\documentstyle[aps,prd,epsfig,eqsecnum]{revtex}
\begin{document}
\draft
\preprint{hep-ph/0210211}
\title{Higgs-boson production associated with a bottom quark at hadron colliders \\with SUSY-QCD corrections}

\author{Junjie Cao $^{a,b}$, Guangping Gao $^b$,  Robert J. Oakes$^c$, Jin Min Yang $^b$ \\\ }

\address{$^a$ CCAST(World Laboratory), P.O.Box 8730, Beijing 100080, China}
\address{$^b$ Institute of Theoretical Physics, Academia Sinica, P.O.Box 2735, Beijing 100080, China}
\address{$^c$ Department of Physics and Astronomy, Northwestern University,
              Evanston, IL 60208, USA}
\date{\today}
\maketitle

\begin{abstract}
The Higgs boson production $p p ~(p\bar p)\to b h +X$ via $b g \to b h$ at
hadron colliders, which may be an important channel for testing the
bottom quark Yukawa coupling, is subject to large supersymmetric
quantum corrections. In this work the one-loop SUSY-QCD
corrections to this process are evaluated and are found to be
quite sizable in some parameter space. We also study the behavior
of the corrections in the limit of heavy SUSY masses and find the
remnant effects of SUSY-QCD. These remnant effects, which are left
over in the Higgs sector by the heavy sparticles, are found to be
so sizable (for a light CP-odd Higgs and large $\tan\beta$) that
they might be observable in the future experiment. The exploration
of such remnant effects is important for probing SUSY, especially
in case that the sparticles are too heavy (above TeV) to be
directly discovered in future experiments.
\end{abstract}

\pacs{14.80.Cp, 13.85.Qk,12.60.Jv}

\section{Introduction}
\label{sec:sec1}
Searching for the Higgs boson is the most
important task for the Fermilab Tevatron collider and the CERN
Large Hadron Collider (LHC).  Among various
Higgs production mechanisms, those induced by the  bottom quark
Yukawa coupling are particularly important because in some
extensions of the Standard Model (SM) such a coupling could be
considerably enhanced and thus the production rates can be much
larger than the SM predictions.
The Minimal Supersymmetric Standard Model (MSSM)\cite{HaberKane}
serve as a good example of such extensions, where the coupling of the
lightest $CP$-even Higgs boson (denoted by $h$) to the bottom quark is
proportional to $\tan{\beta}$ \cite{Gunion} and thus can be
significantly enhanced by large $\tan\beta$.

In the production channels of the Higgs boson via  its  coupling
to the bottom quark , the process $p p~ (p\bar{p} ) \to b h
+X$ via $ b g \to b h$ was recently emphasized in
Ref.\cite{willenbrock}. The advantage of this process over the
production via $b \bar b \to h$ \cite{bbh}, the dominant
production channel via the bottom quark Yukawa coupling, is that
the final bottom quark can be used to reduce backgrounds and to
identify the Higgs boson production mechanism \cite{background}.
And compared with the production via $gg, q \bar q \to h  b
\bar{b}$ \cite{ggbbh}, the production rate of $p p ~(p\bar{p}
) \to b h +X$  is one order of magnitude larger. So the production
$p p ~(p\bar{p} ) \to b h +X$ may be a crucial channel for
testing the bottom quark Yukawa coupling.

If the MSSM is indeed chosen by Nature, then the prediction of the
cross section for the production $p p ~(p\bar{p} ) \to b h
+X$ \cite{huang} must be renewed with the inclusion of SUSY
quantum corrections because, like the process of the charged Higgs
boson production $pp~(p\bar p) \to t H^-+X$ \cite{refree,gao}
and the relevant Higgs decays\cite{decay1,haber-nondecoup,gluino},
the SUSY quantum corrections to this process may be quite large.
In this work we study  the one-loop
SUSY-QCD  corrections to this process, which is believed to
be the dominant part in the SUSY corrections.

It is well known that the low-energy observables in the MSSM
will recover their corresponding SM predictions when all sparticles
as well as $M_A$ (the mass of the CP-odd neutral Higgs boson) take
their heavy limits.
If only some of the masses take their heavy limits, e.g.,  all sparticles
are heavy but $M_A$ is light,  then large remnant effects of SUSY may be
left over in the physical observables of the Higgs sector.
The study of these remnant effects will serve as an important
probe for those heavy SUSY particles \cite{haber-nondecoup}.
Such kind of study will be performed for the process 
$pp ~(p\bar{p} ) \to b h +X$ in this work. After deriving the SUSY-QCD
corrections to this process, we will examine the behavior of
the corrections in the limit of heavy SUSY masses. When  $M_A$
is light and the sparticles take their heavy limits, we find
the large remnant effects left over by the SUSY-QCD in such a Higgs
production process.

This paper is organized as follows. In Section \ref{sec:calculations}
we present our strategy for the calculation of the one-loop SUSY-QCD corrections
to the process $pp ~(p\bar{p} ) \to b h +X$. In
Section \ref{sec:results}, we scan the parameter space of the MSSM to estimate
the size of the SUSY-QCD corrections. In Section \ref{sec:decouple}, we study
the behaviors of these corrections in the limit of heavy
SUSY masses. The conclusion is given in Section  \ref{sec:conclusion} and the
detailed formula obtained in our calculations are presented in the Appendix.

\section{Calculations}
\label{sec:calculations}

At high energy hadron colliders, the incoming $b$-quark is generated from gluons
splitting into nearly collinear $b \bar{b} $ pairs. When one
member of the pair initiates a hard-scattering subprocess, its
partner tends to remain at low $p_T $ and to become part of the
beam remnant. Hence the final state typically has no high-$p_T$
$b$-quarks. When the scale of the hard scattering is large
compared with the $b$-quark mass, the $b$-quark is regarded as part of
the proton sea\cite{bsea}. However, unlike the light quark sea,
the $b$-quark sea is perturbatively calculable. If the scale of the
hard scattering is $\tilde{\mu}$ (for the scale we use $\tilde{\mu}$
to distinguish from the SUSY parameter $\mu$), the $b$-quark distribution
function $b(x, \tilde{\mu}) $ is intrinsically of order $\alpha_s
(\tilde{\mu}) \log(\tilde{\mu}/m_b)$. As $ \tilde{\mu} $
approaches $m_b$ from above, $ b(x, \tilde{\mu}) \rightarrow 0 $;
while as $\tilde{\mu} $ becoming asymptotically large,
$\alpha_s(\tilde{\mu}) \log(\tilde{\mu}/m_b)$ approaches order of
unity and one needs to sum terms of order $\alpha_s^n (\tilde{\mu}) \log^n
(\tilde{\mu}/m_b)$ into the $b$-quark distribution function to yield a
well-behaved perturbation expansion in terms of $\alpha_s $
\cite{bsea}. In this case, the $b$-quark distribution function becomes
of the same order as the light partons.  The main uncertainty of
$b$-quark distribution function comes from that of gluon distribution
function which is about $10\% $ \cite{lai}.

\begin{figure}[hbt]
\begin{center}
\epsfig{file=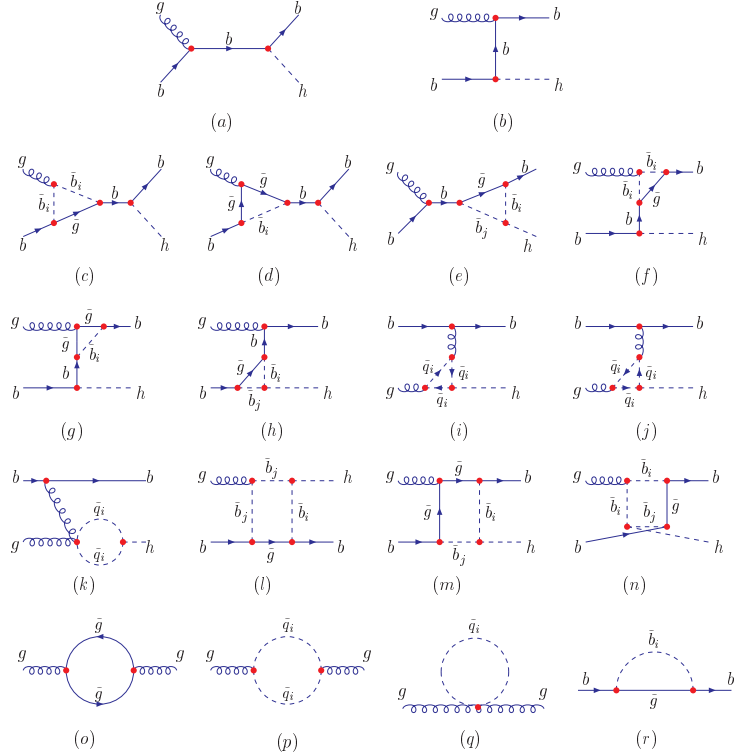,width=10cm} \vspace*{.5cm} \caption{ Feynman
diagrams of $g b \to b h $ with one-loop SUSY-QCD corrections:
$(a,b)$ are tree level diagrams; $(c-e)$ are one-loop vertex
diagrams for $s$-channel; $(f-k)$ are one-loop vertex diagrams for
$t$-channel ; $(l-n)$ are the box diagrams; $(o-r)$ are
self-energy diagrams. } \label{fig:feyman}
\end{center}
\end{figure}

The subprocess  $gb \to b h$ occurs through both $s$-channel and
$t$-channel shown in Fig.~\ref{fig:feyman}$(a, b)$. The spin- and
color-averaged differential cross section at tree-level is given
by
\begin{equation}
\frac{d \hat{\sigma}^{0}}{d \hat{t}} = -
\frac{\alpha_s (\tilde{\mu})}{24} \left (\frac{g m_b
(\tilde{\mu})}{2 m_W} \right )^2 \left(
\frac{\sin{\alpha}}{\cos{\beta}} \right )^2 \frac{1}{\hat{s}^2}
\frac{m_h^4+(\hat{s}+ \hat{t}-m_h^2)^2}{\hat{s} \hat{t}} ,
\label{tree}
\end{equation}
where $\hat{s}$ and $\hat{t}$ are  the usual Mandelstam
variables, $\alpha_s (\tilde{\mu}) $ is the running strong
coupling,  and $m_b (\tilde{\mu}) $ is the running bottom quark
mass\cite{bbh}.  $\alpha$ represents the mixing angel between the
two CP-even Higgs boson eigenstates and $\beta$ is defined by
$\tan\beta=v_2/v_1$ with $v_{1,2}$ denoting the vacuum
expectation values of the two Higgs doublets\cite{Gunion}. In
Eq.(\ref{tree}), we use the $\overline{MS}$ running mass of the 
$b$-quark rather than the pole mass to take into account large QCD
logarithm corrections to the vertex $hb\bar{b}$ \cite{Brat}. The
SM prediction of the cross section is recovered when setting
$|\sin{\alpha}/\cos {\beta}| =1 $\cite{willenbrock}. Throughout
the calculations we neglect the $b$-quark mass except in the $b$-quark
Yukawa couplings.

The one-loop Feynman diagrams of SUSY-QCD corrections are shown in
Fig.~\ref{fig:feyman}(c-r). In our calculations we use dimensional
regularization to control the ultraviolet divergences in the
virtual loop corrections. For the renormalization of strong
coupling constant $g_s$, we employ the $\overline{MS}$ scheme
\cite{Bene}. As to the $ h b \bar{b}$ Yukawa coupling,  at one
loop level to ${\cal O}(\alpha_s)$ it is given by
\begin{eqnarray}
\bar{g}_{hbb}=g_{hbb}+\delta {g}_{hbb}^{QCD}+ \delta {g}_{hbb}^{SQCD} ,
\end{eqnarray}
where $\bar{g}_{hbb}$ denotes the one-loop coupling, $g_{hbb} $ is
tree-level coupling, $\delta{g}_{hbb}^{QCD} $ is the radiative
correction from pure QCD \cite{Brat}, and $\delta{g}_{hbb}^{SQCD}
$ is the one-loop SUSY-QCD contribution \cite{haber-nondecoup}. In
determining $\delta{g}_{hbb}^{QCD} + \delta{g}_{hbb}^{SQCD} $, one
needs the counter-term of the vertex $h b \bar{b} $, whose general
from is given by $ g_{hbb} (\frac{\ \ \ \delta{m}_b^{QCD}}{m_b}+
\frac{\ \ \ \delta{m}_b^{SQCD}}{m_b}) $ with $\delta{m}_b$ being
the counter-term of the $b$-quark mass defined by
$m_b^0=m_b+\delta{m}_b$ ($m_b^0 $ is the bare mass). 
$\delta{m}_b$ is determined by requiring $m_b$ to be the
pole of the one-loop corrected $b$-quark propagator
\cite{Brat,haber-nondecoup,carena}. One major difference between $
\delta{g}_{hbb}^{QCD}$ and $ \delta{g}_{hbb}^{SQCD} $ is that the
former contains large logarithms $ \alpha_s
\log{\frac{\tilde{\mu}}{m_b}} $ of ${\cal{O}} (1)$ 
and thus one needs to introduce $\overline{MS}$ running mass $m_b (\tilde{\mu})
$ to absorb leading logarithms
$\alpha_s^n\log{(\frac{\tilde{\mu}}{m_b})^n }$ \cite{Brat}. An
extensive discussion about this issue in the MSSM was provided in
Ref.\cite{carena}.

The one-loop SUSY-QCD contribution to the amplitude of $g b \to b h$ can be
written as
\begin{eqnarray}
\delta M &=& \frac{i g_{s}^{3}  T^{a} }{16 \pi^2} \frac{g m_b} {2 m_W}
\frac{\sin\alpha}{\cos\beta}
\overline{u} (p_2)  ( C_1 \gamma^{\mu}P_{L} + C_2
\gamma^{\mu}P_{R} + C_3 \gamma^{\mu} \not{k} P_{L} + C_4
\gamma^{\mu} \not{k} P_{R} + C_5 p_1^{\mu} P_{L} + C_6
p_1^{\mu}P_{R}  \nonumber \\
& & + C_7 p_1^{\mu} \not{k} P_{L} + C_8 p_1^{\mu} \not{k} P_{R} +
C_9 p_2^{\mu} P_{L} + C_{10} p_2^{\mu}P_{R} + C_{11} p_2^{\mu}
\not{k} P_{L} + C_{12} p_2^{\mu} \not{k} P_{R}) u (p_1)
\epsilon_{\mu} (k) ,
\end{eqnarray}
where $P_{L,R}\equiv (1\mp \gamma_5)/2$, $T^a\equiv \lambda^a/2$
with  $\lambda^a$ being the Gell-Mann matrices, and $k$, $p_1$ and
$p_2$ are the momentum of the incoming gluon, incoming $b$-quark and the
outgoing $b$-quark, respectively. $g_s$ and $m_b$ should be
understood as the running ones in Eq.(\ref{tree}).
The coefficients $C_{i}$ arise from the loops and are given
explicitly in Appendix A. We have checked that all the ultraviolet
divergences canceled as a result of renormalizability of the
MSSM.

The differential cross section of $g b \to b h$ with one-loop SUSY-QCD corrections
is given by
\begin{eqnarray}
\frac{d \hat{\sigma}}{d \hat{t}} =\frac{d \hat{\sigma}^{0} }{d \hat{t}}
  + \frac{d (\Delta \hat{\sigma})}{d \hat{t}} ,
\end{eqnarray}
where the first term is the tree-level result given in
Eq.(\ref{tree}) and the second term is the one-loop
SUSY-QCD corrections given by
\begin{eqnarray}
\frac{d (\Delta \hat{\sigma})}{d \hat{t}} & = &
-\frac{\alpha_s}{48} \frac{\alpha_s}{4 \pi} \left (
\frac{\sin{\alpha}}{\cos{\beta}} \right )^2  \left (\frac{g m_b}{
2 m_W} \right )^2 \frac{1}{\hat{s}^2} \nonumber \\
& & \times \left [ 2 (C_3+C_4) m_h^2+ (C_5 + C_6)\frac{m_h^2 (\hat{s}+
\hat{t} -m_h^2)}{\hat{t}} + (C_9+ C_{10}) \frac{-(\hat{s}+ \hat{t}
-m_h^2)^2}{\hat{s}} \right ] .  \label{loop}
\end{eqnarray}
The cross section of $g b\rightarrow b h$ is then given by
\begin{equation}
\hat{\sigma}(\hat s) =\int_{\hat{t}_{min}}^{\hat{t}_{max}} {\rm d}
\hat{t} \ \frac{d \hat{\sigma}}{d \hat{t}} ,
\label{cross}
\end{equation}
where $ \hat{t}_{max} = 0 $ and $ \hat{t}_{min}= -\hat{s}+ m_h^2 $
with $m_h$ denoting the Higgs mass. In order to avoid collinear
divergence in Eq.(\ref{cross}) and to enable the outgoing $b$-jet
to be tagged by silicon vertex detector at the Tevatron and the LHC , we
require the transverse momentum of the outgoing $b$-jet to be
larger than $15$ GeV and apply a rapidity cut $ |\eta_b |< 2.5 $
 for the LHC and $ |\eta_b | <2.0 $ for the Tevatron.

The total hadronic cross section for $pp~ (p\bar{p}) \to b
h +X$ can be obtained by folding the subprocess cross section
$\hat{\sigma}$ with the parton luminosity
\begin{equation}
\sigma(s)=\int_{\tau_0}^1 \!d\tau\, \frac{dL}{d\tau}\, \hat\sigma
(\hat s=s\tau) ,  \label{cross1}
\end{equation}
where $\tau_0=m_h^2/s$ and $s$ denotes the $p p ~(p\bar{p})$ 
squared center-of-mass energy. $dL/d\tau$ is the
parton luminosity given by
\begin{equation}
\frac{dL}{d\tau}=\int^1_{\tau} \frac{dx}{x}[f^p_g(x,Q)
f^{p}_b(\tau/x,Q)+(g\leftrightarrow b)] ,
\end{equation}
where $f^p_b$ and $f^p_g$ are the $b$-quark and gluon
distribution functions in a proton, respectively. In our
numerical calculations, we used the CTEQ5L parton distribution
functions~\cite{pm}. We did not distinguish the factorization scale 
$Q$ and the renormalization scale $\tilde{\mu}$, and  assumed 
$\tilde{\mu}=Q=m_h$. 
The scale dependence of our results will be briefly discussed in 
the proceeding section. 

The process $pp ~(p\bar{p}) \to b h +X$ has been extensively
studied \cite{willenbrock} in the framework of the Standard
Model. Its cross section is found to be at the order of $1 $ fb
for the Tevatron and $100 $ fb for the LHC, and  the next-leading-order
(NLO) QCD correction can enhance the production rate by $50\% \sim 60\%$ 
for the Tevatron and $20\%\sim 40\%$ for the LHC \cite{willenbrock}, 
depending on the applied cuts and the Higgs boson mass. 
We will incorporate such QCD corrections in our calculations for
the production rate $\sigma /\sigma_{SM}$.

\section{Numerical results}

\label{sec:results} In this section we will perform a scan over
the SUSY parameter space to show the possible size of the SUSY-QCD
corrections. Before performing numerical calculations, we take a
look at the relevant parameters involved. For the SM parameters,
we took $m_W=80.448$ GeV, $m_Z=91.187$ GeV, $m_t=174.3 $ GeV,
$\bar{m}_b(\bar{m}_b)=4.2$ GeV  \cite{Groom}$, \sin^2 \theta_W
=0.223$ and $\alpha_s(m_Z) = 0.118$ . We used the one-loop QCD
running $\alpha_s(\tilde{\mu})$ and $m_b (\tilde{\mu}) $.

For the SUSY parameters, apart from gluino mass, the mass
parameters of sbottoms are involved. The sbottom squared-mass
matrix is ~\cite{Gunion}
\begin{equation}
M_{\tilde b}^2 =\left(\begin{array}{cc}
m_{{\tilde b}_L}^2& m_b X_b\\
 m_b X_b & m_{{\tilde b}_R}^2
    \end{array} \right),
\end{equation}
where
\begin{eqnarray}
m_{{\tilde b}_L}^2 &=& m_{\tilde Q}^2+m_b^2 + m_Z^2(I_3^b
-Q_b \sin^2\theta_W)\cos(2\beta), \\
m_{{\tilde b}_R}^2 &=& m_{\tilde D}^2+m_b^2
+ m_Z^2 Q_b \sin^2\theta_W\cos(2\beta),\\
X_b&=& A_b-\mu\tan\beta.  \label{smass1}
\end{eqnarray}
Here $m_{\tilde Q}^2$ and $m_{\tilde{D}}^2$ are soft-breaking mass
terms for left-handed squark doublet $\tilde Q$ and right-handed
down squark $\tilde D$, respectively. $A_b $ is the coefficient of the
trilinear term $H_1 \tilde Q \tilde D$  in soft-breaking terms and
$\mu$ the bilinear coupling of the two Higgs doublet in the
superpotential.  $ I_3^b =-1/2 $ and $ Q_b =-1/3 $ are the isospin
and electric charge of the b-quark, respectively. This mass square
matrix can be diagonalized by a unitary rotation
\begin{eqnarray}
  \left(  \begin{array}{c} \tilde b_L \\ \tilde b_R \end{array} \right )
    =\left (
             \begin{array}{cc}
            \cos\theta_b       &-\sin\theta_b\\
           \sin\theta_b       &\cos\theta_b\\
           \end{array} \right )
\left(  \begin{array}{c} \tilde b_1 \\ \tilde b_2 \end{array} \right ),
\end{eqnarray}
and consequently $\theta_b$ and the masses of physical
sbottoms $ \tilde{b}_{1,2} $  can be expressed as
\begin{eqnarray}
\tan{2 \theta_b} &=& \frac{2 m_b X_b}{(m_{\tilde{b}_L}^2
-m_{\tilde{b}_R}^2)}, \label{theta} \\
m_{\tilde{b}_1}^2&= &m_{\tilde{b}_L}^2 \cos^2{\theta_b} +2 m_b X_b
\cos{\theta_b} \sin{\theta_b} + m_{\tilde{b}_R}^2 \sin^2 \theta_b ,\\
m_{\tilde{b}_2}^2&= &m_{\tilde{b}_L}^2 \sin^2{\theta_b} - 2 m_b
X_b \cos{\theta_b} \sin{\theta_b} + m_{\tilde{b}_R}^2
\cos^2{\theta_b} .
 \end{eqnarray}
From Eqs.(\ref{tree},\ref{loop},\ref{cross}) we know that the
cross section also depends on the Higgs mass, $\alpha$ and
$\beta$, which can be determined at tree level by $\tan{\beta}$
and the CP-odd Higgs mass $M_A$ \cite{Gunion}. Noticing the fact
that both the mass and the mixing angle receive large radiative
corrections when the SUSY scale is high above $m_t$ \cite{Haber}, we
used the loop-corrected relations of Higgs masses and mixing angle
\cite{Carena1,Carena2} in the computation of cross section. In our
calculation, we used the program SUBHPOLE2 \cite{Carena1}, where
two-loop leading-log effects of the MSSM are incorporated in the Higgs
masses and the mixing angel, to generate $m_h$ and $\alpha$ needed
for our computation. The input parameters for this program are the
mass parameters in the top sqaurk and sbottom sector, and $M_A$,
$\tan \beta$ and the heavier chargino mass $m_{\tilde{\chi}}$.

We found that the usage of the loop-corrected relations of Higgs
masses and mixing angle is indeed necessary. Comparing with the
results obtained by using tree-level relations for Higgs masses
and mixing angel, the size of SUSY-QCD corrections by using  the
loop-corrected relations is generally magnified from $30\%$ to
$200 \%$\footnote{The main reason for such an enhancement is that  
for a large SUSY scale, the dominant term of the
SUSY-QCD correction is proportional to $\cot\alpha + \tan\beta $ (see
for example, Eqs.(\ref{approx1}, \ref{approx2}, \ref{slow},
\ref{approx4}) in the proceeding section), whose value obtained by using
the loop-corrected
relations of the Higgs masses and the mixing angle is generally larger
than that by using the tree-level relations.}. 
We also checked that this conclusion is
also valid for the SUSY-QCD correction to the Higgs partial width $\Gamma
(h \to b \bar{b})$.

Now we know the relevant parameters are
\begin{equation} \label{para}
m_{\tilde{Q}}, m_{\tilde{U}}, m_{\tilde{D}}, A_{t,b},
m_{\tilde{g}}, m_{\tilde{\chi}}, \mu, M_A, \tan{\beta} ,
\label{input}
\end{equation}
where $M^2_{\tilde U}$ is the soft-breaking mass term for
right-handed top-squark and  $A_t$ the coefficient of the
soft-breaking trilinear term $H_2 \tilde Q \tilde U$. To
show the main features of SUSY effects in  $ pp ~(p\bar{p})
\to b h +X$, we performed a scan over this ten-dimensional
parameter space. In our scan we make no assumptions about the
relations among these parameters to keep our result
model-independent, but restrict the parameters with mass dimension
to be less than $2$ TeV. In addition, we consider the following
experimental constraints:
\begin{itemize}
\item[{\rm(1)}]
   $\mu>0$ and $\tan\beta$ in the range $5\le\tan\beta\le 50$, which seems
   to be favored by the muon $g-2$ measurement~\cite{Brown01}.
\item[{\rm(2)}]
   The LEP and CDF lower mass bounds on Higgs, gluino, stop, sbottom and chargino~\cite{LEP,PDG00}
\begin{eqnarray}
m_h \geq 114~GeV, m_{{\tilde t}_1}\geq 86.4~GeV,~ m_{{\tilde
b}_1}\geq 75.0~GeV, ~ m_{\tilde{g}}\geq 190~GeV, ~
m_{\tilde{\chi}}^{light} \geq 67.7~GeV ,\label{constrain}
\end{eqnarray}
\end{itemize}
where $m_{\tilde{\chi}}^{light} $ is the mass of the lighter chargino.

\vspace*{-0.7cm}
\begin{figure}
\begin{center}
\epsfig{file=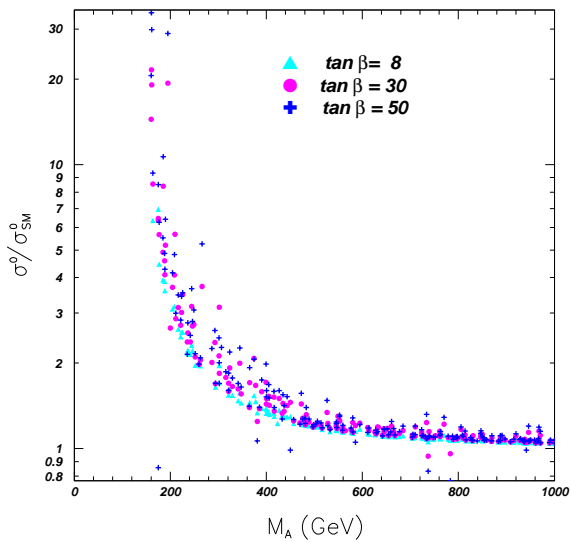,width=8cm} \caption{The scatter plot of
$\sigma^0/\sigma_{SM}^{0}$ versus $M_A$.} 
\label{fig:scan1}
\end{center}
\end{figure}

It would be interesting to first scan over the allowed parameter
space to figure out how large the production rate is enhanced in
the MSSM. In Fig.\ref{fig:scan1} we present the tree-level cross
section relative to the SM prediction with the same Higgs mass.
This ratio is independent of collider energy, but is dependent on
the SUSY mass parameters since we use the loop-corrected
relations of the Higgs masses and the mixing angle (see the second
paragraph of Sect. II) . From Fig.\ref{fig:scan1} one sees that
the production rate in the MSSM can be significant larger than the
SM prediction for a light $M_A $; while for a heavy $M_A$ the MSSM
prediction approaches to the SM value. This character was first
noticed in \cite{haber-decoup} and, as a result of this character,
distinguishing the lightest MSSM Higgs boson  from the SM Higgs
boson in the large $M_A$ limit will be very difficult. When
SUSY-QCD corrections are added, this character remains unchanged
for heavy sbottoms(see following discussions). From
Fig.\ref{fig:scan1} one also finds that there exists the
possibility (although very rare) that the MSSM cross section is
suppressed to be below the SM value \cite{Gordon}. In this case,
the SUSY-QCD corrections will play a more important role in Higgs
phenomenology at colliders \cite{Carena3}.

\begin{figure}
\begin{center}
\epsfig{file=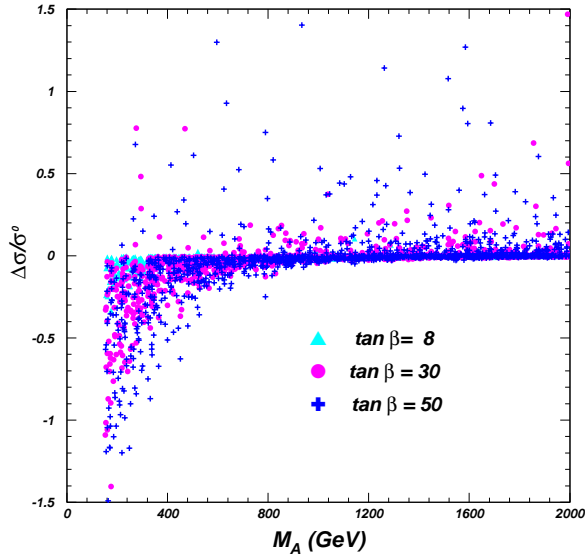,width=8cm} \caption{The scatter plot of the
SUSY-QCD correction $\Delta \sigma/ \sigma^0$ versus $M_A$ for the Tevatron.
The difference of the results between the LHC and the Tevatron
is at the level of parts per mill. }
\label{fig:scan2}
\end{center}
\end{figure}

Now let us scan over the allowed parameter space to show the
possible size of the SUSY-QCD corrections relative to the
tree-level value. In our numerical evaluation, we found the
relative correction is insensitive to collider energy. 
The difference of the results between the LHC and the Tevatron
is at the level of parts per mill. 
In Fig.\ref{fig:scan2}
we present the SUSY-QCD corrections to the cross section.
\vspace*{-0.7cm}
 Fig.\ref{fig:scan2} manifests three features of
SUSY-QCD corrections for large $\tan{\beta}$. The first one is that the correction size
is enhanced by $\tan{\beta}$ and thus can be quite large. The second one is that
for $M_A$ lighter than $500$ GeV, the correction tends to be negative.
The third one is that for large $M_A$, the correction may be positive
and the maximum value seems to be independent
of the value of $M_A$. These features can be explained as follows.

For the correction size larger than $ 2 \% $,  the dominant
contribution of the SUSY-QCD correction is from the loop
corrections to the vertex $h b \bar{b}$ and the contribution of
the box diagrams is much smaller for the parameters satisfying the
constraints in the paragraph following Eq.(\ref{input}) \footnote{In the large limit of
SUSY mass parameters discussed in the proceeding section, we have checked
that, even for the correction size far smaller than $1 \% $,
the dominant contribution still comes from the corrections to the
vertex $ h b \bar{b} $.}. As a result, the correction behaves like
(which is similar to the SUSY-QCD correction to the vertex $h b
\bar{b}$ in case of heavy sbottoms \cite{haber-nondecoup})
\begin{eqnarray}
\frac{\Delta \sigma}{\sigma^0} \sim C_1 \frac{M_{EW}^2}{M_A^2} +C_2
\frac{M_{EW}^2}{M_{\tilde{b}}^2} , \label{explain}
\end{eqnarray}
where  $M_{EW}$ and $M_{\tilde{b}} $ denote the electroweak scale
and the typical mass of sbottoms, respectively.  $C_{1} $ and $C_2 $
are functions of $m_{\tilde{b}_1}$, $m_{\tilde{b}_2}$,
$m_{\tilde{g}}$, $\mu $ and $A_b$, but independent of $M_A$.  It
is found that in general $C_1$ is negative and $C_2 $ is positive
and either $C_1$ or both $C_1$ and $C_2$ are enhanced by large
$\tan{\beta}$. For a light $M_A$ compared with $m_{\tilde{b}}$,
the first term of the RHS in Eq.(\ref{explain}) is dominant and
hence the cross section tends to be negative. While for a large
$M_A$ , the second term is dominant and the cross section tends to
be positive. So the behavior of Eq.(\ref{explain}) can explain the
features of Fig.\ref{fig:scan2}.

From Fig.\ref{fig:scan2} we noticed  that in some corners
of parameter space the one-loop SUSY-QCD contributions to the cross section
are comparable or even larger than the tree-level result and
consequently, one must consider higher order corrections. In
such cases, it is important to sum over the terms $\alpha_s^n
(\frac{\mu}{M_{SUSY}})^n $ to all orders of perturbation theory by
using an effective Lagrangian approach \cite{carena,technique}.
\begin{figure}[htb]
\begin{center}
\epsfig{file=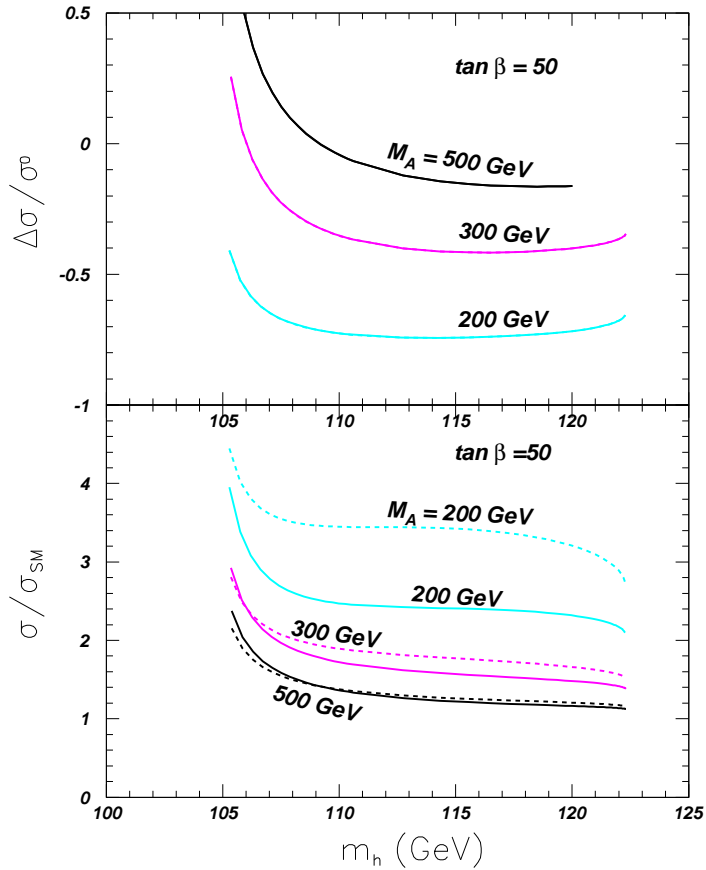,width=9.7cm} \caption{The
SUSY-QCD correction $\delta \sigma/\sigma^0 $ and the cross
section $\sigma /\sigma_{SM} $ versus the mass of the produced Higgs boson.
Solid curves are for the LHC and the dashed for the Tevatron. 
For $\Delta \sigma/ \sigma^0$, each solid curve overlaps with the 
corresponding dashed one due to the tiny difference.}
\label{fig:higgs}
\end{center}
\end{figure}
Next we study the dependence of the SUSY-QCD correction $\delta
\sigma/\sigma^0 $ and the cross section normalized by the SM
prediction, $\sigma /\sigma_{SM} $, on the mass of the produced
Higgs boson, which can be directly compared with experiment
results and hence is much informative. 
In such a study we assumed
a common value ($M_{SUSY}$) for all input SUSY mass parameters
and, considering the fact that $m_h$ is 
insensitive to $M_A$ for $M_A >150 GeV$ \cite{Carena2}, we fixed
the value of $M_A$. Then through varying the value of $M_{SUSY}$,
we obtain the different mass value of the Higgs boson. 

The dependence on $m_h$ is illustrated in
Fig.\ref{fig:higgs} for $\tan\beta =50$. 
(Note that $m_h$  can vary only in a small range since it is 
   stringently upper bounded in the MSSM.) 
In this figure and also in the
following figures showing $\sigma /\sigma_{SM}$, we also
incorporated the conventional QCD corrections \cite{willenbrock}
into both the MSSM and the SM cross sections. As pointed out
earlier, the SUSY-QCD correction $\delta \sigma/\sigma^0$ is not 
sensitive to collider energy. The difference of the results between
the LHC and the Tevatron is too small to be visible, as shown
in the upper part of Fig.\ref{fig:higgs}. But for 
$\sigma /\sigma_{SM}$ the difference  between
the LHC and the Tevatron is visible since the QCD corrections
are significantly different for these two colliders \cite{willenbrock}. 

We also studied the dependence of the production rate on the renormalization scale
$\tilde \mu$. (As pointed out earlier, we assume that the factorization scale 
is equal to the renormalization scale.)
We found that such a scale dependence is significant 
in some parameter space. For example, for the LHC with 
$M_A=300$ GeV and $m_h=120 $ GeV, the ratio $\sigma(\tilde \mu)/\sigma (m_h)$
is $0.93$ for $\tilde \mu=m_h/2$ and $1.03$ for $\tilde \mu=2 m_h$. 
Such an uncertainty is comparable with the  uncertainty from the $b$-quark Yukawa 
coupling ($ \bar{m}_b (\bar{m}_b) = 4.2 \pm 0.2 $) and the partron
distribution function (about $10 \%$).

\section{Behaviours of SUSY-QCD corrections in decoupling limits}
\label{sec:decouple}
To study the behaviors of the SUSY-QCD correction in
the large limit of SUSY mass parameters, we consider four typical
cases as in Ref.\cite{haber-nondecoup} where the decoupling
property of SUSY-QCD correction to the coupling of $h b \bar{b} $ is
analyzed.
To qualitatively understand the feature of each case, we present
the approximate formula in the limits, but in practical numerical
calculations we use the complete one-loop expressions.

About the inputs of the SUSY parameters, there are several
differences between our work and Ref.\cite{haber-nondecoup}. The
first one is that in \cite{haber-nondecoup} the tree-level
relations for the Higgs masses and the mixing angle were used, but in our
calculations we use the loop-corrected relations.
As discussed earlier, using the
loop-corrected relations leads to a significantly different
correction. The second one is that in our analysis we considered
the experimental bounds in Eq.(\ref{constrain}). This will rule
out some parameter space which have been considered in
Ref.\cite{haber-nondecoup}. The third one is that in cases B and D
we also require $A_{t,b}$ to be large since large $A_{t,b}$ is
favored by the Higgs mass bound.

\begin{figure}[htb]
\begin{center}
\epsfig{file=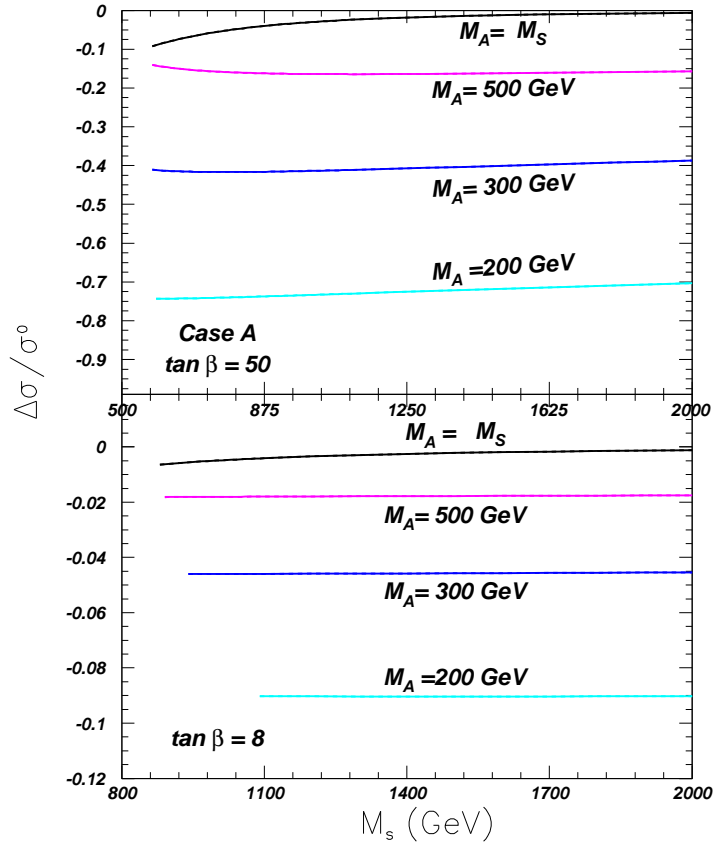,width=9.7cm} \caption{The SUSY-QCD correction
$\Delta \sigma/ \sigma^0$ versus $M_S$ in Case A. For each fixed 
value of $M_A$,  the solid curve (for the LHC) overlaps with the
corresponding dashed one (for the Tevatron) due to the tiny 
difference.} \label{fig:cas1A}
\end{center}
\end{figure}

(1) {\em Case A:}~~ All SUSY mass parameters except $M_A$ are of
the same size (collectively denoted by $M_S$) and tend to heavy,
i.e.,
\begin{eqnarray}
m_{\tilde Q} \sim m_{\tilde U} \sim m_{\tilde D} \sim A_b \sim A_t
\sim m_{\tilde g}\sim \mu  \sim M_S .
\end{eqnarray}
In this case the SUSY-QCD correction behaves like
\begin{eqnarray}
\frac{\Delta \sigma}{\sigma^0} \sim \frac{2 \alpha_s}{3 \pi} \left
[ - ( \tan{\beta}+\cot{\alpha}) -\cot{\alpha} (\frac{m_h^2}{12
M_S^2}+ \frac{m_b^2 \tan^2{\beta}}{2 M_S^2}) +
\frac{\tan{\beta}}{3} \frac{\cos{\beta} \sin(\alpha
+\beta)}{\sin{\alpha}} \frac{m_Z^2}{M_S^2} \right ] ,
\label{approx1}
\end{eqnarray}
where the first term in the RHS corresponds to the first term in
Eq.(\ref{explain})\footnote{In the MSSM the tree-level relation
for Higgs masses and mixing angle predicts the following relation:
$\cot{\alpha} = -\tan{\beta} -\frac{2 m_Z^2}{M_A^2} \tan{\beta}
\cos{2 \beta} +O(\frac{m_Z^4}{M_A^4}) $ and at loop level, the gap
between $\cot{\alpha} $ and $-\tan{\beta} $ is generally
enlarged.}  and the rest corresponds to the second term in
Eq.(\ref{explain}). The striking feature of this case is that for
very large $ M_S $, the correction approaches a nonzero constant,
and this remnant effect of SUSY-QCD corrections is enhanced by
large $\tan{\beta} $. This feature is illustrated in
Fig.\ref{fig:cas1A}. From Fig.\ref{fig:cas1A} one also finds that
the SUSY-QCD correction in this case is negative and sizable
\footnote{If the correction is too sizable (say exceed 50\%),
higher order corrections are also important and need a proper
treatment\cite{carena,technique}.} for a light $ M_A $ and a large
$\tan{\beta}$.

\begin{figure}[htb]
\begin{center}
\epsfig{file=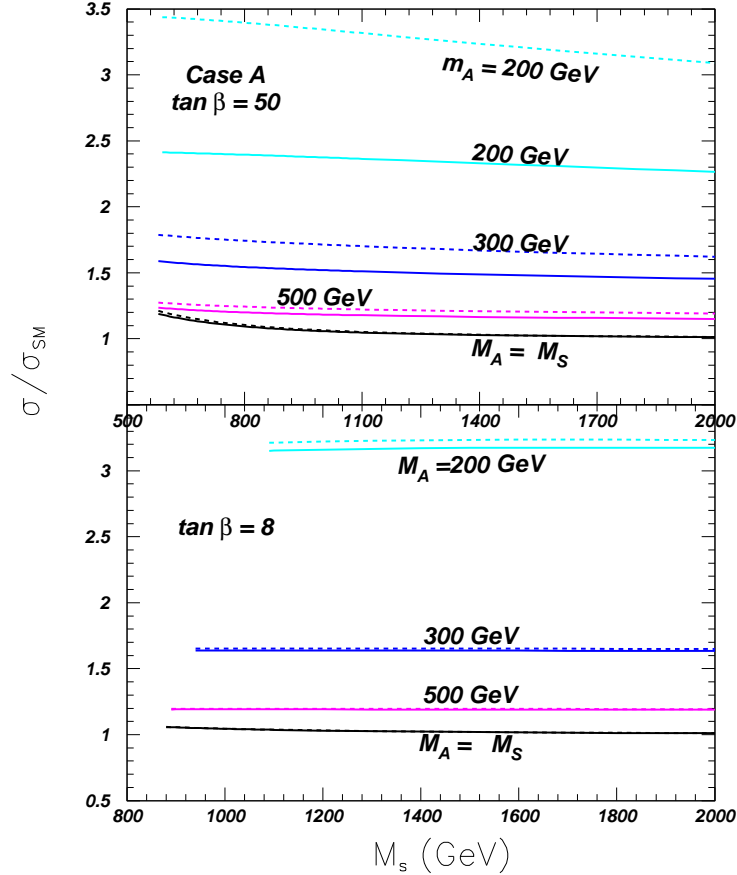,width=10cm} 
\caption{ $\sigma/\sigma_{SM}$ versus $M_A $ in Case A
          for the LHC (solid) and the Tevatron (dashed). }
\label{fig:acas1A}
\end{center}
\end{figure}

\begin{figure}[htb]
\begin{center}
\epsfig{file=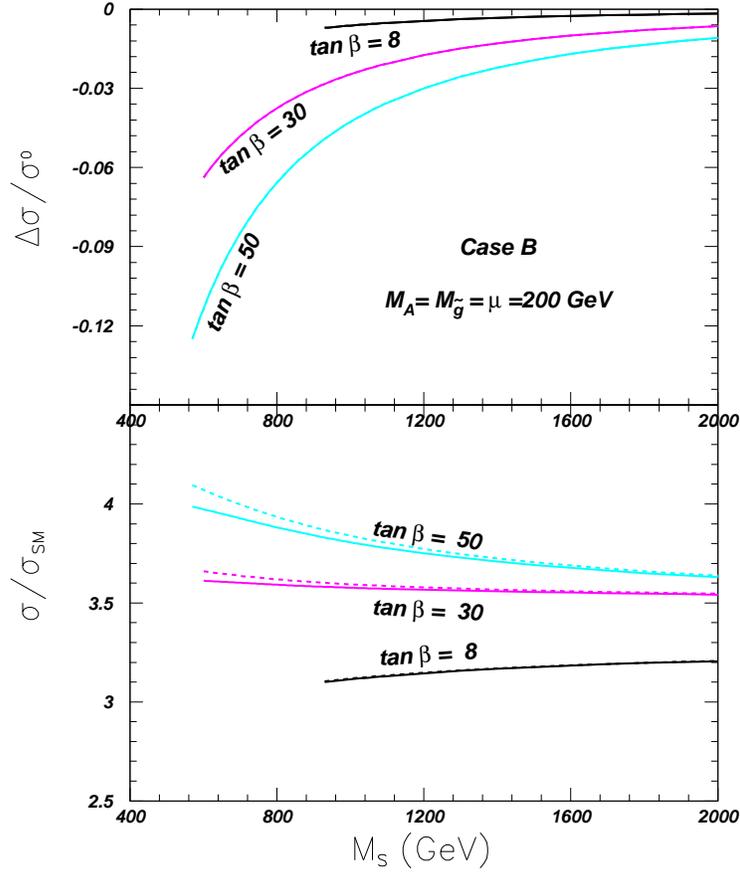,width=10cm} \caption{ The
SUSY-QCD correction $\Delta \sigma/ \sigma^0$ and
$\sigma/\sigma_{SM}$ versus $M_S$ in Case B. The solid curves are for 
the LHC and the dashed for the Tevatron. 
For $\Delta \sigma/ \sigma^0$, each solid curve overlaps with the 
corresponding dashed one due to the tiny difference. } \label{fig:casB}
\end{center}
\end{figure}

\begin{figure}[htb]
\begin{center}
\epsfig{file=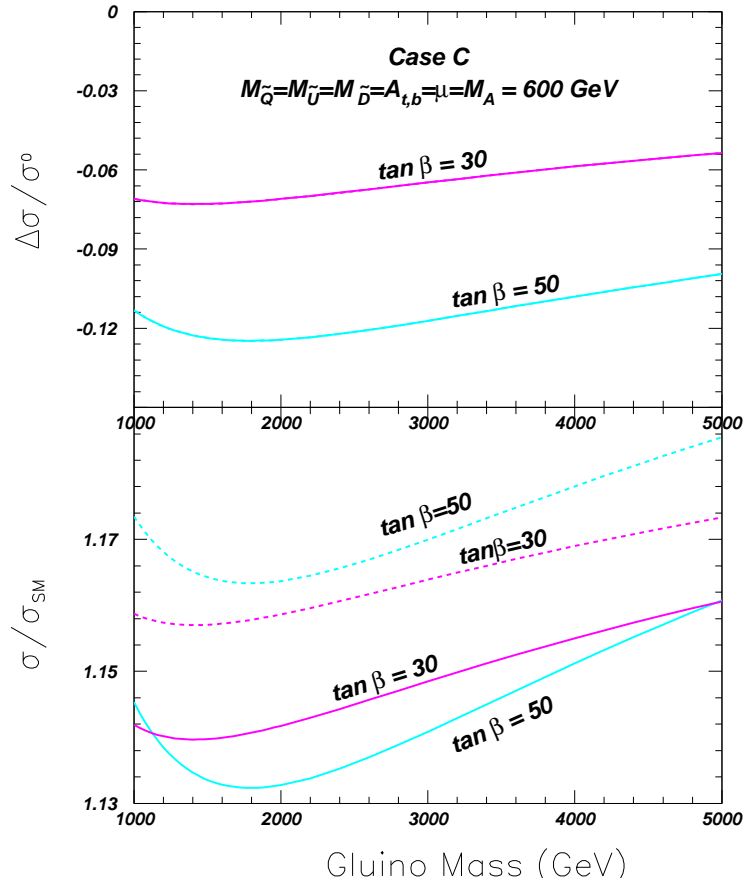,width=10cm} \caption{ The
SUSY-QCD correction $\Delta \sigma/ \sigma^0$ and
$\sigma/\sigma_{SM}$ versus the gluino mass in Case C. 
The solid curves are for the LHC and the dashed for the Tevatron. 
For $\Delta \sigma/ \sigma^0$, each solid curve overlaps with the 
corresponding dashed one due to the tiny difference.} \label{fig:casC}
\end{center}
\end{figure}

In Fig.\ref{fig:acas1A} we show the loop corrected cross section
normalized by the SM prediction.  From this figure we see that for
$M_A$ of several hundred GeV, although the tree-level cross
section in the MSSM can be reduced by large SUSY-QCD corrections,
an enhancement of several times over the SM prediction can still
be expected due to the fact that  the tree-level $b$-quark Yukawa
coupling in the MSSM is significantly enhanced by large
$\tan\beta$ for light $M_A $.  This large enhancement shows a very
weak dependence on $M_S$. So we can conclude that up to the
next-leading order, a light $M_A$ is still able to make the MSSM
cross section larger than the SM prediction.
(2) {\em Case B:}~~
    $M_{\tilde Q}$ $M_{\tilde U}$, $M_{\tilde D}$ and $A_{t, b} $
(collectively denoted as $M_S $) is much larger than $\mu $,
$m_{\tilde{g}} $ and $M_A $ (collectively denoted as $M $ ), i.e.,
\begin{eqnarray}
M_{\tilde Q, \tilde U, \tilde D} \sim A_{t, b}  \sim  M_S  \gg
m_{\tilde g} \sim \mu \sim M_A \sim M .
\end{eqnarray}
In this case the SUSY-QCD correction behaves as
\begin{eqnarray}
\frac{\Delta \sigma}{\sigma^0}
&\sim & \frac{2 \alpha_s}{3 \pi} \left [ \frac{-2 M^2}{M_S^2} (\tan{\beta}+\cot{\alpha} )
- \frac{m_h^2 M^2}{ 6 M_S^4} (\frac{M_S}{M}+\cot{\alpha}) \right . \nonumber \\
& & \left . -\frac{m_Z^2}{2 M_S^2} \frac{\cos{\beta}
\sin{(\alpha+\beta)}}{\sin{\alpha}} (1-
(\frac{M_S}{M}-\tan{\beta}) \frac{2 M^2 }{M_S^2} ) \right ].
\label{approx2}
\end{eqnarray}
From this expression we see that in the large $M_S$ limit, the
SUSY-QCD corrections decouple rapidly as $M^2/M_S^2 $ and the
decoupling behavior is slowed down by large $\tan{\beta} $. The
characters of this case are shown in Fig.\ref{fig:casB}.
So we see that even with a fixed light $M_A$, the process still
does not have remnant SUSY-QCD effects if the gluino mass and $\mu$
are also kept light.
(3) {\em Case C:}~~ Only the gluino mass gets  much larger than other
SUSY parameters (collectively denoted as $M_S$) \footnote{In this case
the Higgs mass bound requires $M_S $ to be much
larger than electroweak scale.}:
\begin{eqnarray}
m_{\tilde{g}} \gg M_{\tilde Q, \tilde U, \tilde D} \sim A_{t, b}
\sim \mu \sim M_A \sim M_S .
\end{eqnarray}
In this case the SUSY-QCD correction behaves as
\begin{eqnarray}
\frac{\Delta \sigma}{\sigma^0}
&\simeq & \frac{2 \alpha_s}{3 \pi} \left [
\frac{2 M_S}{M_{\tilde{g}}} (\tan{\beta}+ \cot{\alpha})
(1- \log{\frac{M_{\tilde{g}}^2}{M_S^2}} ) -
\frac{M_S \cot{\alpha}}{3 M_{\tilde{g}}} \frac{m_h^2}{M_S^2} \right .
\nonumber \\
&& \left . +
\frac{M_S \tan{\beta}}{M_{\tilde{g}}} \frac{m_Z^2}{M_S^2}
\frac{\cos{\beta} \sin{(\alpha+\beta)}}{\sin{\alpha}} -
\frac{m_b^2 \tan^2{\beta} \cot{\alpha}}{M_{\tilde{g}} M_S} \right ] .
\label{slow}
\end{eqnarray}
The main character of this case is that as gluino mass gets large,
the correction drops very slowly like $\frac{1}{m_{\tilde{g}}}
\log{\frac{m_{\tilde{g}}^2}{M_S^2}}$, which was also observed in
Refs.\cite{haber-nondecoup,gluino}. Again, like other cases, the
size of the correction is enhanced by large $\tan{\beta}$.
In Fig.\ref{fig:casC}  we show the dependence of the SUSY-QCD correction
and $\sigma/\sigma_{SM}$ on the gluino mass. Note that in this case
we found that $\tan{\beta}=8$ cannot satisfy the experimental
bounds in Eq.(\ref{constrain}) for $M_S =600$ GeV. In
Fig.\ref{fig:casC} the correction size is
significantly smaller than those in Case A and B. The reason is that
here a large $M_A$ is chosen so that $ \tan{\beta}+ \cos\alpha $
is suppressed (see footnote 3).

Let us explain the origin of the slowness of the decoupling in
this case. Such slowness of the decoupling arises from the first
term in Eq.(\ref{slow}), i.e., $\frac{2 \mu }{M_{\tilde{g}}}
(\tan{\beta}+ \cot{\alpha}) \log{\frac{M_{\tilde{g}}^2}{M_{\tilde
q}^2}}$ (note that the $M_S$ in the factor $\frac{2 M_S
}{M_{\tilde{g}}}$ is $\mu$ and the one in the logarithm is squark
mass $M_{\tilde q}$). As $M_{\tilde{g}}$ gets much larger than
$M_{\tilde q}$ and $\mu$, $\frac{2 \mu}{M_{\tilde{g}}}$ decreases
but $\log{\frac{M_{\tilde{g}}^2}{M_{\tilde q}^2}}$ increases. For
the example shown in Fig.\ref{fig:casC}, i.e.,  $M_{\tilde{g}}$ is
changing from 1 TeV to 5 TeV with fixed  $M_{\tilde q}=\mu=600$
GeV,  the factor $\frac{2 \mu}{M_{\tilde{g}}}$ is decreased by
$1/5$ but the factor $\log{\frac{M_{\tilde{g}}^2}{M_{\tilde
q}^2}}$ is increased by 4.16. Thus the slowness of the decoupling
as gluino gets heavy is caused by the enlarged mass splitting
between gluino and squark. Of course,  since $\tan{\beta}+
\cot{\alpha}$ is proportional to $\frac{2 m_Z^2}{M_A^2}
\tan{\beta} \cos{2 \beta}$ (see footnote 3), the contribution of
the first term in Eq.(\ref{slow}) will be decoupled rapidly if
$M_A$ gets large.

(4) {\em Case D:} ~~One of the sbottoms and $A_{t,b} $ become
heavy while other mass parameters (denoted as M)  are fixed. We choose
\begin{eqnarray}
M_{\tilde D} \sim M_{\tilde U} \sim A_{t,b} \gg M_{\tilde Q} \sim m_{\tilde{g}}
\sim \mu \sim M_A \sim M \gg M_{EW}
\end{eqnarray}
or equally
\begin{eqnarray}
m_{\tilde{b}_2} \sim m_{\tilde{t}_2} \sim A_{t,b} \gg
m_{\tilde{b}_1} \sim m_{\tilde{t}_1} \sim m_{\tilde{g}} \sim \mu
\sim M \gg M_{EW} ,
\end{eqnarray}
where $ m_{\tilde{b}_{1,2}}$ and $ m_{\tilde{t}_{1,2}} $
are the masses of bottom-squarks and top-squarks, respectively.

\begin{figure}[htb]
\begin{center}
\epsfig{file=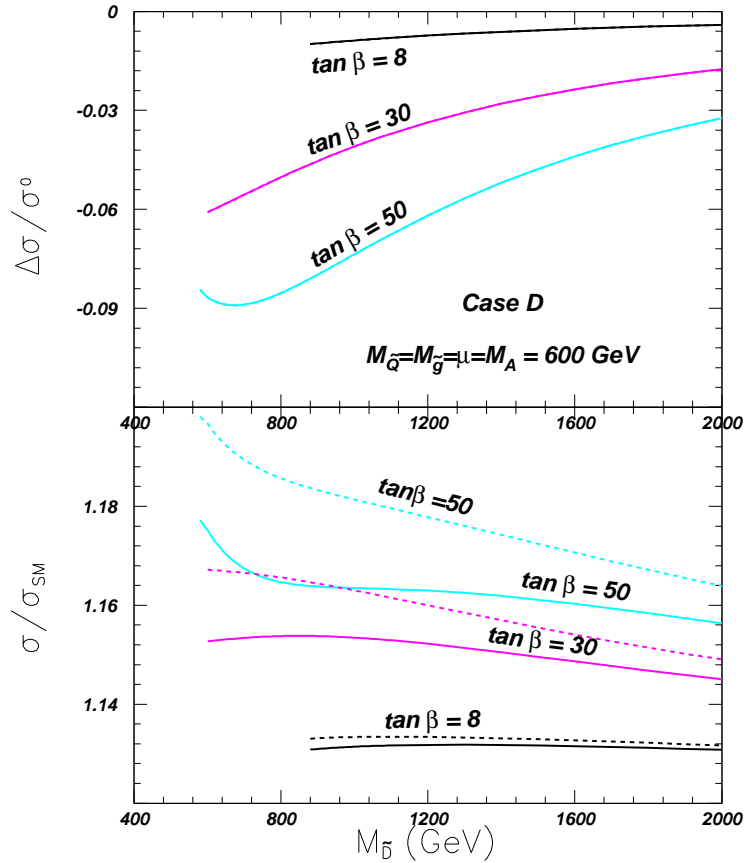,width=10cm} \caption{ The
SUSY-QCD correction $\Delta \sigma/ \sigma^0$ and
$\sigma/\sigma_{SM}$ versus $M_{\tilde D}$ in Case D for the LHC (solid
curves) and the Tevatron (dashed curves). 
For $\Delta \sigma/ \sigma^0$, each solid curve overlaps with the 
corresponding dashed one due to the tiny difference.} \label{fig:casD}
\end{center}
\end{figure}

In this case the SUSY-QCD correction behaves as
\begin{eqnarray}
\frac{\Delta \sigma}{\sigma^0} & \simeq & \frac{2 \alpha_s}{3 \pi}
\left [ \frac{2 M^2}{m_{\tilde{b}_2}^2} (\tan{\beta}+
\cot{\alpha}) \left (1+ \log\frac{M^2}{m_{\tilde{b}_2}^2} \right )
+ \frac{m_Z^2}{m_{\tilde{b}_2}^2} \frac{\cos{\beta}
\sin{(\alpha+\beta)}}{\sin{\alpha}} (-1+\frac{2}{3} s_W^2)
(\frac{m_{\tilde{b}_2}}{m_{\tilde{b}_1}} -\tan{\beta} ) \right ].
\label{approx4}
\end{eqnarray}
The main feature of this case is the correction decouples like $
\frac{1}{m_{\tilde{b}_2}^2} \log\frac{M^2}{m_{\tilde{b}_2}^2} $
and this decoupling behavior is slowed down by large $ \tan{\beta} $.
Fig.\ref{fig:casD} shows the dependence of the correction and
the cross section on $m_{\tilde D}$ ($\sim m_{\tilde{b}_2}$).
Comparing with the results in case B, we see that
the correction size decrease more slowly as $m_{\tilde{D}}$
gets heavy.

From the above analyses we see that when $M_A$ is fixed and all
other SUSY mass parameters get large, the SUSY-QCD lefts over some
remnant effects in the Higgs production process $ p p \to b h +X $
at the hadron colliders. Note that for the remnant effects to be
left over, $\mu$ and gluino mass must be comparable with or larger
than the masses of the sbottoms. The fundamental reason for such
a behavior is that the couplings like $h\tilde{b}_i \tilde{b}_j$
are proportional to SUSY mass parameters.

We conclude this section by  making  a few remarks. Firstly,
in our analysis we assumed $m_{\tilde{g}}, \mu > 0$ and, as a result, all four cases
have negative values of the correction. In the
anomaly-mediated SUSY breaking scenario\cite{anomal}, a negative
$m_{\tilde{g}}$ is predicted and in this case, the sign of
the correction may be reversed.  Secondly, it should be noted that in
the calculations of such Higgs processes it is necessary to use the
loop-corrected relations of the Higgs masses and the mixing angle since such
relations can significantly affect the results.
Thirdly, in case of a light $ M_A$, although the
SUSY-QCD corrections tend to reduce the cross section severely,
the MSSM cross section can still be several
times larger than the SM prediction. If the cross section of
this process is measured in the future and found to be
several times larger than the SM prediction, a light $M_A $ is
favored.

\section{Conclusion}
\label{sec:conclusion}

In this work, we studied the one-loop SUSY-QCD quantum effects in
the Higgs production $ p p ~(p\bar p) \to b h +X $ at the Tevatron and the
LHC in the framework of the MSSM. We found that for a light $M_A$
and large $\tan{\beta}$, the corrections can be quite sizable and
cannot be neglected. We performed a detailed analysis on the
behaviors of the corrections in the limits of heavy SUSY masses
and found that when $M_A$ is fixed and all other SUSY mass
parameters get large, the SUSY-QCD lefts over some remnant effects
in the Higgs production process $ p p \to b h +X $. Such remnant
effects can be so sizable for a light $M_A$  that they might be observable in the future
experiment. The exploration of such remnant effects is important
for probing SUSY, especially in case that the sparticles are too
heavy (above TeV) to be directly discovered in  future
experiments.

\section*{Acknowledgment}
We thank Tao Han for discussions.
This work is supported in part by the Chinese Natural Science Foundation
and by the US Department of Energy, Division of High Energy Physics
under grant No. DE-FG02-91-ER4086.
\appendix
\section{Expressions of form factors}

Before presenting the explicit form of $C_i$s, we define the
following abbreviations:
\begin{eqnarray}
\hat{s}&=& (p_1+k)^2,\ \ \ \ \ \ \hat{t}= (k-p_2)^2,     \\
a_{1,2}&=&\frac{1}{\sqrt{2}} (\sin{\theta_b} \mp \cos{\theta_b}),\
\ \ \ \ \ \ \ \ \ b_{1,2}=\frac{1}{\sqrt{2}} (\cos{\theta_b} \pm
\sin{\theta_b}),    \\
A_{I}^L&=& (a_I-b_I)^2, \ \ \ \ A_{I}^R=(a_I + b_I)^2,\ \ \ \
A_I=a_I^2-b_I^2,\ \ \ \ (I=1,2),   \\
A_{IJ}^L&=& a_I a_J+b_I b_J+ a_I b_J+ b_I a_J, \ \ \ \ \
A_{IJ}^R=a_I
a_J+b_I b_J- a_I b_J - b_I a_J,     \\
B_{IJ}^L&=&a_I a_J-b_I b_J - a_I b_J+ b_I a_J, \ \ \ \ \
B_{IJ}^R=a_I a_J - b_I b_J + a_I b_J - b_I a_J,    \\
cc_1 &=& -\frac{m_Z}{\cos{\theta_W}} \left (\frac{1}{2}
-\frac{1}{3} \sin^2{\theta_W} \right ) \sin(\alpha+\beta),
  \\ cc_2& = & -\frac{m_Z}{\cos{\theta_W}} \frac{1}{3}
\sin^2{\theta_W} \sin(\alpha+\beta), \ \ \ \
cc_3 = \frac{m_b}{2 m_W \cos{\beta}} (A_b \sin{\alpha} + \mu \cos{\alpha}),    \\
Q_{11}&=&cc1 \cos^2\theta_b +cc2 \sin^2 \theta_b +2 cc3
\sin\theta_b \cos \theta_b,   \\
Q_{12}&=&(cc2-cc1) \sin{\theta_b}\cos{\theta_b}
+ cc3 (\cos^2\theta_b -\sin^2\theta_b),     \\
Q_{21}&=&(cc2-cc1) \sin{\theta_b}\cos{\theta_b} + cc3
(\cos^2\theta_b -\sin^2\theta_b),   \\
Q_{22}&= &cc1 \sin^2\theta_b +cc2 \cos^2 \theta_b - 2 cc3
\sin\theta_b \cos \theta_b,     \\
B^{I}_i&= & B_i (p, m_{\tilde{g}}, m_{\tilde{b}_I}) |_{p^2=m_b^2},\\
B^{s\ I}_i&=& B_i (p, m_{\tilde{g}}, m_{\tilde{b}_I})|_{p^2=\hat{s}},
\ \ \ \ \ \ B^{t\ I}_i = B_i (p, m_{\tilde{g}},
m_{\tilde{b}_I}) |_{p^2=\hat{t}},   \\
C^{a\ I}_{ij}&=&C_{ij}(p_1,k,m_{\tilde{b}_I}, m_{\tilde{g}},
m_{\tilde{g}}), \ \ \ \ C^{b\ I}_{ij}
=C_{ij}(-p_1,-k,m_{\tilde{g}}, m_{\tilde{b}_I},
m_{\tilde{b}_I} ),   \\
C^{c\ I}_{ij}&=&C_{ij}(-p_2,k,m_{\tilde{b}_I}, m_{\tilde{g}},
m_{\tilde{g}}), \ \ \ \ C^{d\ I}_{ij} =C_{ij}(-p_2,k,m_{\tilde{g}},
m_{\tilde{b}_I}, m_{\tilde{b}_I} ),   \\
C^{e\ IJ}_{ij}&=&C_{ij}(-p_2,-p_h, m_{\tilde{g}}, m_{\tilde{b}_I},
m_{\tilde{b}_J}), \ \ \ \ C^{f\ IJ}_{ij} =C_{ij}(-p_1,p_h,
m_{\tilde{g}}, m_{\tilde{b}_J}, m_{\tilde{b}_I} ),   \\
D^{g\ IJ}_{ij}&=&D_{ij} (-p_1,-k, p_h, m_{\tilde{g}},
m_{\tilde{b}_J}, m_{\tilde{b}_J}, m_{\tilde{b}_I}), \ \ \ \ D^{h\
IJ}_{ij}=D_{ij} (-p_1, p_h, p_2, m_{\tilde{g}}, m_{\tilde{b}_J},
m_{\tilde{b}_I}, m_{\tilde{g}}), \\
D^{k \ IJ}_{ij}&=&D_{ij} (-p_2, k, -p_h, m_{\tilde{g}},
m_{\tilde{b}_I}, m_{\tilde{b}_I}, m_{\tilde{b}_J}),
\end{eqnarray}
where $B_i$, $C_{ij} $ and $D_{ij} $ are loop functions defined in
\cite{hooft}.

After $ \frac{g_s^2}{16 \pi^2} $ is factored out, the
renormalization constant of b-quark can be expressed as
\begin{eqnarray}
\delta Z_L & = & C_F \sum_{I=1}^2 A_I^L B_1^I,\ \ \ \ \ \delta Z_R
 =  C_F \sum_{I=1}^2 A_I^R B_1^I,     \\
\frac{\delta m_b}{m_b}& = & C_F \sum_{I=1}^2
(\frac{m_{\tilde{g}}}{m_b} A_I B_0^I -\frac{1}{2} A_I^L B_1^I
-\frac{1}{2} A_I^R B_1^I) ,
\end{eqnarray}
where $C_F=4/3 $. The contributions of the self-energy diagrams of $b$-quark propagator
can be written as
\begin{eqnarray}
\Sigma_L^s&=&C_F \sum_{I=1}^2 A_I^L B_1^{s\ I} - \delta Z_L, \ \ \
\ \Sigma_L^t=C_F \sum_{I=1}^2 A_I^L B_1^{t\ I} - \delta Z_L, \\
\Sigma_R^s&=&C_F \sum_{I=1}^2 A_I^R B_1^{s\ I} - \delta Z_R, \ \ \
\ \Sigma_R^t=C_F \sum_{I=1}^2 A_I^R B_1^{t\ I} - \delta Z_R.
\end{eqnarray}

$C_i$ appeared in Eq.(\ref{loop}) are given by
\begin{eqnarray}
C_3 &= & \sum_{I=1}^2 \{ -\frac{3}{2} (\hat{s} C_{12}^{a\ I} +
\hat{s} C_{23}^{a\ I}+ 2 C_{24}^{a\ I} -1/2 - m_{\tilde{g}}^2
C_{0}^{a\ I}) A_I^L/\hat{s}+ \frac{1}{3}  C_{24}^{b\ I} A_I^L/\hat{s} \nonumber  \\
& &  -\frac{3}{2} (\hat{t} C_{12}^{c\ I} + \hat{t} C_{23}^{c\ I}+
2 C_{24}^{c\ I} -1/2 - m_{\tilde{g}}^2 C_{0}^{c\ I})
A_I^R/\hat{t}+ \frac{1}{3}  C_{24}^{d\ I} A_I^R /\hat{t} \} \nonumber\\
&& -\delta Z_L/\hat{s}-\delta
Z_R/\hat{t}-\Sigma_L^s/\hat{s}-\Sigma_R^t/\hat{t}-(\frac{1}{2}
\delta Z_L+\frac{1}{2} \delta Z_R +\frac{\delta m_b}{m_b} )
(\frac{1}{\hat{s}}+\frac{1}{\hat{t}}) \nonumber  \\
&&+ \frac{2 m_W \cos{\beta}}{m_b \sin{\alpha}} \sum_{I,J=1}^2
Q_{IJ} \{ \frac{4}{3} m_{\tilde{g}} C_{0}^{e\ IJ} B_{IJ}^L
/\hat{s}+\frac{4}{3} m_{\tilde{g}} C_{0}^{f\ IJ} B_{IJ}^L
/\hat{t}+ \frac{3}{2} m_{\tilde{g}} D_{0}^{h\ IJ} B_{IJ}^L \},\\
C_4 &= & \sum_{I=1}^2 \{ -\frac{3}{2} (\hat{s} C_{12}^{a\ I} +
\hat{s} C_{23}^{a\ I}+ 2 C_{24}^{a\ I} -1/2 - m_{\tilde{g}}^2
C_{0}^{a\ I}) A_I^R/\hat{s}+ \frac{1}{3}  C_{24}^{b\ I} A_I^R/\hat{s}\nonumber   \\
& &  -\frac{3}{2} (\hat{t} C_{12}^{c\ I} + \hat{t} C_{23}^{c\ I}+
2 C_{24}^{c\ I} -1/2 - m_{\tilde{g}}^2 C_{0}^{c\ I})
A_I^L/\hat{t}+ \frac{1}{3}  C_{24}^{d\ I} A_I^L /\hat{t} \} \nonumber\\
&& -\delta Z_R/\hat{s}-\delta
Z_L/\hat{t}-\Sigma_R^s/\hat{s}-\Sigma_L^t/\hat{t}-(\frac{1}{2}
\delta Z_L+\frac{1}{2} \delta Z_R +\frac{\delta m_b}{m_b} )
(\frac{1}{\hat{s}}+\frac{1}{\hat{t}})   \nonumber \\
&&+ \frac{2 m_W \cos{\beta}}{m_b \sin{\alpha}} \sum_{I,J=1}^2
Q_{IJ} \{ \frac{4}{3} m_{\tilde{g}} C_{0}^{e\ IJ} B_{IJ}^R
/\hat{s}+\frac{4}{3} m_{\tilde{g}} C_{0}^{f\ IJ} B_{IJ}^R
/\hat{t}+ \frac{3}{2} m_{\tilde{g}} D_{0}^{h\ IJ} B_{IJ}^R \}, \\
C_5 &= & \sum_{I=1}^2 \{ -\frac{3}{2} (-4 C_{24}^{a\ I} +1 + 2
m_{\tilde{g}}^2 C_{0}^{a\ I}) A_I^L/\hat{s}- \frac{1}{3} ( \hat{s}
C_{12}^{b\ I}+\hat{s} C_{23}^{b\ I}+ 2 C_{24}^{b\ I} )  A_I^L/\hat{s} \} \nonumber  \\
&& + 2 \delta Z_L/\hat{s}+ 2 \Sigma_L^s/\hat{s}+( \delta Z_L+
\delta Z_R +2 \frac{\delta m_b}{m_b} )/\hat{s}+ \frac{2 m_W
\cos{\beta}}{m_b \sin{\alpha}} \sum_{I,J=1}^2 Q_{IJ} \{
-\frac{8}{3} m_{\tilde{g}} C_{0}^{e\ IJ} B_{IJ}^L /\hat{s} \nonumber \\
&& +\frac{1}{3} m_{\tilde{g}} (D_{0}^{g\ IJ}+
D_{11}^{g\ IJ}-D_{13}^{g\ IJ} ) B_{IJ}^L + 3 m_{\tilde{g}}
(D_{11}^{h\ IJ}- D_{12}^{h\ IJ}) B_{IJ}^L +\frac{1}{3} m_{\tilde{g}} D_{13}^k B_{IJ}^L \},\\
C_6 &= & \sum_{I=1}^2 \{ -\frac{3}{2} (-4 C_{24}^{a\ I} +1 + 2
m_{\tilde{g}}^2 C_{0}^{a\ I}) A_I^R/\hat{s}- \frac{1}{3} ( \hat{s}
C_{12}^{b\ I}+\hat{s} C_{23}^{b\ I}+ 2 C_{24}^{b\ I} )  A_I^R/\hat{s} \} \nonumber  \\
&& + 2 \delta Z_R/\hat{s}+ 2 \Sigma_R^s/\hat{s}+( \delta Z_L+
\delta Z_R +2 \frac{\delta m_b}{m_b} )/\hat{s}+ \frac{2 m_W
\cos{\beta}}{m_b \sin{\alpha}} \sum_{I,J=1}^2 Q_{IJ} \{
-\frac{8}{3} m_{\tilde{g}} C_{0}^{e\ IJ} B_{IJ}^R /\hat{s} \nonumber\\
&& +\frac{1}{3} m_{\tilde{g}} (D_{0}^{g\ IJ}+ D_{11}^{g\
IJ}-D_{13}^{g\ IJ} ) B_{IJ}^R + 3 m_{\tilde{g}} (D_{11}^{h\ IJ}-
D_{12}^{h\ IJ}) B_{IJ}^R +\frac{1}{3} m_{\tilde{g}} D_{13}^k
B_{IJ}^R \},
  \\
C_9 &= & \sum_{I=1}^2 \{ -\frac{3}{2} (-4 C_{24}^{c\ I} +1 + 2
m_{\tilde{g}}^2 C_{0}^{c\ I}) A_I^R/\hat{t}- \frac{1}{3} ( \hat{t}
C_{12}^{d\ I}+\hat{t} C_{23}^{d\ I}+ 2 C_{24}^{d\ I} )  A_I^R
/\hat{t} \} \nonumber  \\
&& + 2 \delta Z_R/\hat{t}+ 2 \Sigma_R^t/\hat{t}+( \delta Z_L+
\delta Z_R +2 \frac{\delta m_b}{m_b} )/\hat{t}+ \frac{2 m_W
\cos{\beta}}{m_b \sin{\alpha}} \sum_{I,J=1}^2 Q_{IJ} \{
-\frac{8}{3} m_{\tilde{g}} C_{0}^{f\ IJ} B_{IJ}^L /\hat{t} \nonumber\\
&& +\frac{1}{3} m_{\tilde{g}} D_{13}^{g\ IJ} B_{IJ}^L
+ 3 m_{\tilde{g}} (D_{12}^{h\ IJ}- D_{13}^{h\ IJ}) B_{IJ}^L +\frac{1}{3}
m_{\tilde{g}} (D^k_0+D^k_{11}-D^k_{13}) B_{ij}^L \}, \\
C_{10} &= & \sum_{I=1}^2 \{ -\frac{3}{2} (-4 C_{24}^{c\ I} +1 + 2
m_{\tilde{g}}^2 C_{0}^{c\ I}) A_I^L/\hat{t}- \frac{1}{3} ( \hat{t}
C_{12}^{d\ I}+\hat{t} C_{23}^{d\ I}+ 2 C_{24}^{d\ I} )  A_I^L
/\hat{t} \}  \nonumber \\
&& + 2 \delta Z_L/\hat{t}+ 2 \Sigma_L^t/\hat{t}+( \delta Z_L+
\delta Z_R +2 \frac{\delta m_b}{m_b} )/\hat{t}+ \frac{2 m_W
\cos{\beta}}{m_b \sin{\alpha}} \sum_{I,J=1}^2 Q_{IJ} \{
-\frac{8}{3} m_{\tilde{g}} C_{0}^{f\ IJ} B_{IJ}^R /\hat{t}
\nonumber \\ && +\frac{1}{3} m_{\tilde{g}} D_{13}^{g\ IJ} B_{IJ}^R
+ 3 m_{\tilde{g}} (D_{12}^{h\ IJ}- D_{13}^{h\ IJ}) B_{IJ}^R
+\frac{1}{3} m_{\tilde{g}} (D^k_0+D^k_{11}-D^k_{13}) B_{ij}^R  \}.
\end{eqnarray}

Since we have neglect the b-quark mass throughout this paper,
$C_{1,2,7,8,11,12}$ are irrelevant to our result and we do not
present their explicit forms here.


\begingroup\raggedright\endgroup

\end{document}